\documentclass[%
 reprint,
 superscriptaddress,
 amsmath,amssymb,
aps,
prb,
floatfix,
]{revtex4-1}

\usepackage{graphicx}% Include figure files
\usepackage{dcolumn}% Align table columns on decimal point
\usepackage{bm}% bold math
\usepackage{braket} %para escribir bra-kets
\usepackage[caption=false]{subfig} % <=== "caption=false" added
\usepackage{changepage}
\usepackage{amsmath}
\usepackage{xcolor}
\usepackage{xfrac}
\usepackage{multirow}
\usepackage{appendix}
\usepackage{soul}
\usepackage[version=4]{mhchem}

\def\gup{$\Gamma_1^+$}
\def\gum{$\Gamma_1^-$}
\def\gdp{$\Gamma_2^+$}
\def\gdm{$\Gamma_2^-$}

\def\gcp{$\Gamma_4^+$}
\def\gcm{$\Gamma_4^-$}
\def\gsp{$\bar{\Gamma}^+$}
\def\gsm{$\bar{\Gamma}^-$}

\def\yup{$U_1U_4{}^+$}
\def\yum{$U_1U_4{}^-$}
\def\ydp{$U_2U_3{}^+$}
\def\ydm{$U_2U_3{}^-$}
\def\ysp{$\bar{U}^+$}
\def\ysm{$\bar{U}^-$}

\definecolor{amethyst}{rgb}{0.6, 0.4, 0.8}

\begin{document}

\title{Emergent heavy-fermion physics in a new family of topological insulators  RAsS (R = Y, La, and Sm)}

\author{Iñigo Robredo}
\email{inigo.robredo@list.lu}
\affiliation{Luxembourg Institute of Science and Technology (LIST), Avenue des Hauts-Fourneaux 5, L-4362 Esch/Alzette, Luxembourg}
\affiliation{Donostia International Physics Center, 20018 Donostia-San Sebastián, Spain}

\author{Yuan Fang}
\affiliation{Department of Physics \& Astronomy,  Extreme Quantum Materials Alliance, Smalley-Curl Institute,
Rice University, Houston, Texas 77005, USA}

\author{Lei Chen}
\affiliation{Department of Physics \& Astronomy,  Extreme Quantum Materials Alliance, Smalley-Curl Institute,
Rice University, Houston, Texas 77005, USA}
\affiliation{Department of Physics and Astronomy, Stony Brook University, Stony Brook, NY 11794, USA}

\author{Nazar Zaremba}
\author{Yurii Prots}
\author{Mitja Krnel}
\author{Markus König}
\affiliation{Max Planck Institute for Chemical Physics of Solids, Dresden 01187, Germany} \relax

\author{Thomas Doert}
\affiliation{Technical University of Dresden, Dresden 01062, Germany}

\author{Jeroen van den Brink}
\affiliation{Institute for Theoretical Solid State Physics, IFW Dresden, Helmholtzstrasse 20, 01069 Dresden, Germany}

\author{Claudia Felser}
\affiliation{Max Planck Institute for Chemical Physics of Solids, Dresden 01187, Germany}

\author{Qimiao Si}
\email{qmsi@rice.edu}
\affiliation{Department of Physics \& Astronomy,  Extreme Quantum Materials Alliance, Smalley-Curl Institute,
Rice University, Houston, Texas 77005, USA}

\author{Eteri Svanidze}
\email{svanidze@cpfs.mpg.de}
\affiliation{Max Planck Institute for Chemical Physics of Solids, Dresden 01187, Germany}

\author{Maia G. Vergniory}
\email{maia.vergniory@usherbrooke.ca}
\affiliation{Donostia International Physics Center, 20018 Donostia-San Sebastián, Spain}
\affiliation{D\'epartement de Physique et Institut Quantique,  
Universit\'e de Sherbrooke, Sherbrooke, J1K 2R1, Qu\'ebec, Canada.}
\affiliation{Regroupement Qu\'eb\'ecois sur les Mat\'eriaux de Pointe (RQMP), Quebec H3T 3J7, Canada}

\begin{abstract}
Realizing topological phases in strongly correlated materials has become a major impetus in condensed matter physics. Although many compounds are now classified as topological insulators, $f$‑electron systems—with their strong electron correlations—provide an especially fertile platform for emergent heavy‑fermion phenomena driven by the interplay of topology and many‑body effects. In this study, we examine the crystalline topology of a new RAsS series (R = Y, La, Sm), revealing a structural variant from previous reports. We demonstrate that YAsS and SmAsS host hourglass fermions protected by glide symmetry. SmAsS notably exhibits a strong effective‐mass enhancement, placing it alongside SmB${}_6$ and YbB${}_{12}$ as a material that couples topological surface states with emergent Kondo physics, yet distinguished by its crystalline symmetry constraints and $f$–$p$ orbital hybridization. To capture these features, we construct a minimal model incorporating $f$‑electron degrees of freedom, which reproduces the observed topological properties and predicts that the surface states survive in the correlated regime, albeit shifted in energy. Our work thus introduces a new family of correlated topological materials and forecasts the robustness of their surface states under Kondo correlations.

\end{abstract}

\maketitle

\section{Introduction}

\begin{figure*}
    \centering
    \includegraphics[width=\linewidth]{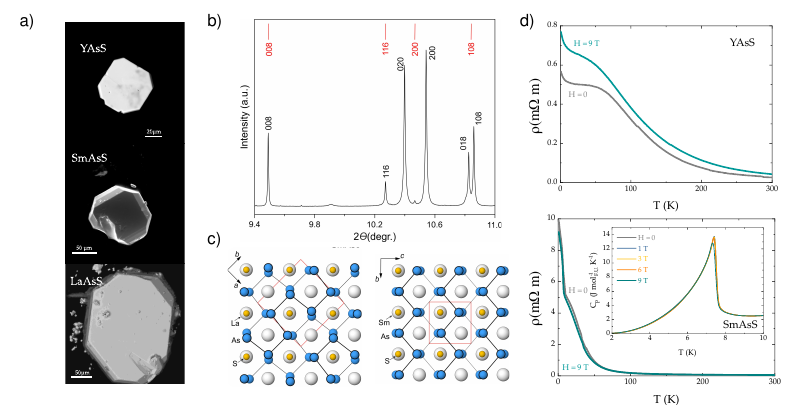}
    \caption{a) Scanning electron micrographs of RAsS compounds. b) Selected region diffraction pattern of SmAsS collected by using synchrotron radiation ($\lambda$ = 0.35466 \AA) at beamline ID22 at ESRF (for more details, see text). c) Crystal structure of LaAsS (left) and SmAsS (right) projected along the largest (pseudo-tetragonal) translation period. The location of the zigzag and trapezoidal chains of As atoms differs by $\frac{1}{2}$ of viewing direction, which is emphasized by thin and thick lines, respectively. d) Resistivity measurements for YAsS and SmAsS. Both compounds display semiconductor-like resistivity. For the SmAsS compound, entrance into the antiferromagnetic state is marked by a sharp anomaly around $T = 7.5$ K.}
    \label{fig:exp}
\end{figure*}

The discovery of topological materials has reshaped our understanding of quantum phases, offering potential breakthroughs in next-generation technologies, from fault-tolerant quantum computing to dissipationless electronic devices \cite{doi:10.1126/science.1231473,Aghaee2025,doi:10.1126/sciadv.ado4756}. However, despite the rapid progress in classifying and engineering weakly correlated topological systems \cite{TQC,MTQC,alltopobands,completecatalogue,magneticht}, a crucial missing piece in this landscape is the realization of topological phases in strongly correlated materials. The combination of nontrivial band topology with electronic correlations is predicted to give rise to exotic quasiparticles, interaction-driven topological transitions, and even novel forms of superconductivity \cite{Checkelsky2024,Hu2024,PhysRevB.101.245159} —yet experimentally confirmed examples remain scarce. Heavy-fermion systems, where flat bands from localized $f$-electrons hybridize with itinerant conduction states, provide a fertile ground for realizing such physics, as their many-body interactions can dynamically modify topological properties in ways that go beyond conventional band theory \cite{Tokura2022}. Notable examples are Kondo systems, both Kondo insulators such as SmB${}_6$, YbB${}_{12}$ \cite{iraola2023topology,Chang2017,Sundermann2015,PhysRevB.90.201106,Hagiwara2016} and 
Weyl-Kondo semimetals \cite{lai2018weyl,PhysRevB.101.075138,PhysRevLett.118.246601} such as realized in Ce$_3$Pd$_3$Bi$_4$ \cite{PhysRevLett.118.246601,doi:10.1073/pnas.2013386118}.
However, heavy-fermion topological crystalline insulators (TCIs)—whose nontrivial topology is enforced by crystalline symmetries remain rare.

In this work, we report the discovery of two new topological crystalline insulators with hourglass fermions, namely SmAsS and YAsS. Furthermore, the SmAsS compound is a heavy-fermion candidate and LaAsS is close to a topological transition. These materials are closely related to the ZrSiS aristotype family of square-net semimetals, which have recently been in the spotlight of the topological condensed matter community in regard to their topological properties \cite{Leslie1,Leslie2,Leslie3,Leslie4}. At the same time, we resolve a long-standing misclassification of the three compounds that were previously reported with inaccurate crystallographic structures \cite{CEOLIN1977137}. Combining single crystal technique with high-resolution powder diffraction performed by means of a synchrotron source, we accurately establish their atomic arrangements, which differ significantly from prior reports. 
Using the formalism of topological quantum chemistry (TQC)\cite{completecatalogue,alltopobands}, we determine the topological invariants of these compounds and show that two of them host glide-symmetry protected Hourglass fermions (Y/SmAsS) while the other is close to a topological transition (LaAsS). Using DFT calculations and effective tight-binding models (TB), and 
incorporating the correlated $f$-electrons,
we are able to derive the simplest model that captures the fundamental topological properties of the three systems. Interestingly, this model serves as the physical realization of the layer construction used to explain similar topological properties in related systems, such as LaSbTe \cite{PhysRevB.92.205310,Qian2020} or ErAsS \cite{https://doi.org/10.1002/adma.202110664}. The measurements of electrical resistivity of YAsS and SmAsS single crystals reveal semiconducting behavior for both, while a large value of the Sommerfeld coefficient $\gamma = 160$ mJ mol$^{-1}_\text{Sm}$ K$^{-2}$ for SmAsS signals significant effective electron mass enhancement.

\begin{figure*}
    \centering
    \includegraphics[width=\linewidth]{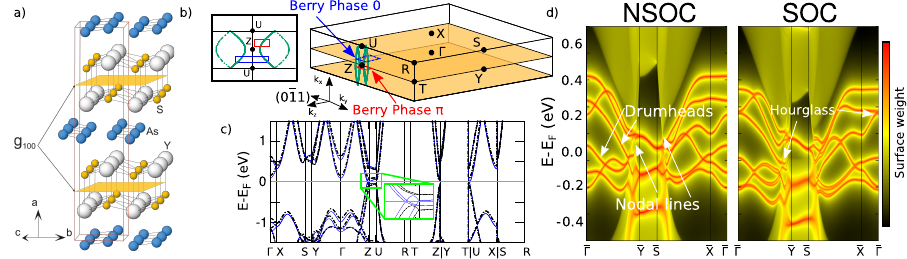}
    \caption{Bulk and surface properties of YAsS. a) Crystal structure of YAsS in $Pnma$ space group (\#62), with the glide symmetry responsible for hourglass fermions displayed in orange. b) BZ of orthorhombic $P$ lattice. In orange, the plane in reciprocal space that is invariant under the action of the glide symmetry. In  green, the nodal lines enforced by the $z_2'=1$ topological invariant in the SOC-free phase. They are very close to each other, in neighboring BZs. We show the nodal lines in a projection parallel to $k_y$ in the inset. In the Cartesian axis we show the reciprocal space direction in which hourglass states are predicted to exist in the surface. c) Bulk band structure of YAsS without and with SOC. The inset shows the nodal line crossings in the absence of SOC. d) Calculated (0-11) surface spectrum of YAsS. In the left, spinless surface spectrum. The white arrows point at the surface projection of bulk nodal lines, from which surface drumhead states emerge. In the right, spinful surface spectrum, showing the hourglass fermions.}
    \label{fig:fig1}
\end{figure*}

This work is divided as follows: In Sec.~\ref{sec:sec1} we present the new crystal structures of the family of compounds under study. In Sections \ref{sec:YAsS}, \ref{sec:SmAsS} and \ref{sec:LaAsS} we study each compound's electronic and topological properties in depth.
We present our conclusions in Sec.~\ref{sec:conc}. In Sec.~\ref{sec:methods} we provide the details on the numerical calculations. Finally, we show further details on the model and the theory of Hourglass fermions in Appendices~\ref{app:app1} and \ref{app:app2}, respectively, as well as full structural details in Appendix~\ref{app:app3}.

\section{Crystal structure and magnetic behavior}\label{sec:sec1}

The crystal structure models of rare earth arsenic sulfides RAsS reported previously as well as results of present publication show different kind of deformations starting from the tetragonal aristotype ZrSiS \cite{Onken1964}, realized e.g. for UAsS, space group $P4/nmm$, $a = 3.879$ \AA, $c = 8.168$ \AA \cite{Hulliger1968}. For this reason, we use in this manuscript non-standard settings for the derivatives of the tetragonal aristotype to better illustrate the relationships between different crystal structure models and to discuss the particularities encountered in crystal structure determination. 

Initially, a monoclinic structure was established for CeAsS (reported: space group $P112_1/b$, $a = 4.047$ \AA, $b = 5.616$ \AA, $c = 17.45$ \AA, $\gamma = 135.85^{\circ}$, pseudo-tetragonal setting: space group $P 1 1 21/n$, $a = 4.047$ \AA, $b = 3.912$ \AA, $c = 17.45$ \AA, $\gamma = 90.26^{\circ}$) \cite{Societe1907} and later assumed for the  whole RAsS series (with the exception of Eu and Yb), based on the powder X-ray diffraction data \cite{CEOLIN1977137}. Thus, transformation of the lattice reported for SmAsS ($a = 3.93$ \AA, $b = 5.48$ \AA, $c = 17.00$ \AA, $\gamma = 135.45^{\circ}$) into pseudo-tetragonal setting also resulted in the monoclinic angle close to 90°: $a = 3.93$ \AA, $b = 3.84$ \AA, $c = 17.00$ \AA, $\gamma = 90.37^{\circ}$. Later, the crystal structure of SmAsS was resolved as orthorhombic on the basis of single crystal data (space group $Pcmn$, $a = 3.90$ \AA, $b = 3.85$ \AA, $c = 17.11$ \AA) \cite{Strauss}.
% Based on these structures, recent investigations using the TQC formalism \cite{TQC,completecatalogue,alltopobands} determined that both YAsS and LaAsS are TCIs with symmetry indicator (SI) $\mathbb{Z}_4=2$.
Our re-examination of the crystal structure of the representatives of this series has shown that the laboratory equipment -- either by powder diffraction or by single crystal experiments -- is insufficient for a clear and precise identification of the respective atomic arrangement due to the non-negligible deformation of the unit cell. The task was made even more difficult by the tendency of the analyzed specimens to form twinned agglomerates. For this reason, high-resolution X-ray diffraction with synchrotron radiation (beamline ID22 at the ESRF) was needed in order to clearly identify the crystal structure. Based on this data, it was determined that the reported assignment of the monoclinic structure to the entire series is not accurate and a re-examination is required for each individual phase (see Fig.~\ref{fig:exp}a and Fig.~\ref{fig:exp}b).

Thus, the YAsS, SmAsS and LaAsS structures discussed here differ from those reported more than forty years ago \cite{CEOLIN1977137}. YAsS (SmAsS) and LaAsS show different types of deformation, starting from the tetragonal aristotype ZrSiS. While an orthorhombic distortion along the original tetragonal axes $a$ and $b$ is observed for Y/Sm (space group $Pcmn$, $a = 3.913 $ \AA, $b = 3.861$ \AA, $c = 17.146$ \AA, non-standard setting of the space group $Pnma$), a deformation along the diagonals is detected for the La phase (space group $Pmnb$, $a = 5.692$ \AA, $b = 5.707$ \AA, $c = 17.553$ \AA). As a result, the respective lattice parameters differ by a factor of $\sqrt{2}$. The different deformation of the original tetragonal unit cell is due to the different arrangement of the As atoms in the structures. While the As atoms in the SmAsS-type structure are connected to form zigzag chains, the As chains in the LaAsS structure have a trapezoidal arrangement and are stretched in a $45^{\circ}$ direction compared to the location of the As chains in the SmAsS structure, see Fig.~\ref{fig:exp}c. This kind of chain has been frequently referred to as a cis-trans chain \cite{Demchyna2004,Czulucki2010}. All these unit cell deformations are confirmed by the splitting of the corresponding reflections on the high-resolution X-ray powder diffractograms. In Fig.~\ref{fig:exp}b we show a selected region of the diffraction pattern of SmAsS collected by using synchrotron radiation ($\lambda$ = 0.35466 \AA). The indexes of the experimental pattern correspond to the non-standard setting ($Pcmn$) of the space group $Pnma$, $a = 3.913 $ \AA, $b = 3.861$ \AA, $c = 17.146$ \AA. The bars in the upper part of Fig.~\ref{fig:exp}b correspond to the theoretically calculated positions of the tetragonal subcell with the lattice parameter $a = 3.887$ \AA, averaged from the respective values $a$ and $b$ of the real structure. This figure clearly shows the presence of the split reflections, which is typical for the orthorhombic distortion of the unit cell (h0l), and the absence of the additional reflection splitting of the hhl series, which would be expected in the case of the monoclinic distortion. For reference, the 00l reflections remain unsplit in both cases. In the case of LaAsS (space group $Pmnb$, $a = 5.692$ \AA, $b = 5.707 $ \AA, $c = 17.553$ \AA) with the orthorhombic deformation along the $ab$ diagonal, the reflections of the hhl series are split, while the h0l reflexes remain unsplit. The difference in the deformation of the tetragonal unit cell of the ZrSiS aristotype, which led to two reported structural models, can be illustrated using the group-subgroup relationship (for a schematic representation, see Bärnighausen tree, shown in Fig.~\ref{fig:BT}). Note that the calculations haven been done using the standard $Pnma$ setting of the space group types. We show full structural details of the three compounds in Supplementary Tables \ref{T1}, \ref{T2} and \ref{T3} in Appendix~\ref{app:app3}.

We also performed temperature- and field-dependent resistivity measurements of YAsS and SmAsS (see Fig.~\ref{fig:exp}d). In the normal state of SmAsS, we find a large value of the Sommerfeld coefficient $\gamma = 160$ mJ mol$^{-1}_\text{Sm}$ K$^{-2}$, which signals effective electron mass enhancement. This is consistent with having flat bands at the Fermi level from very localized $f$-electrons from Sm atoms, which can lead to heavy-fermion physics in this compound. 

\begin{figure*}
    \centering
    \includegraphics[width=\linewidth]{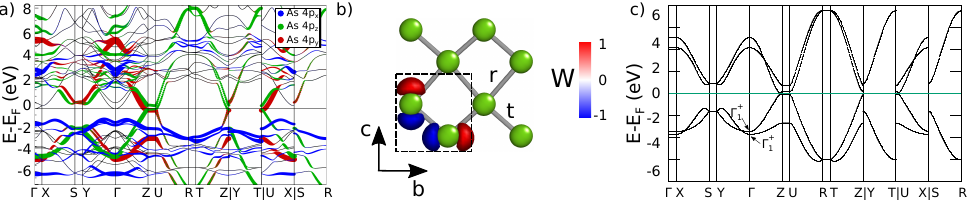}
    \caption{a) Spinless bulk bands of YAsS, with orbital weights from $p_x$, $p_y$ and $p_z$ orbitals of As. b) Wannier functions from $p_z$ and $p_y$ orbitals at As sites in one layer. The color-scale shows the sign of the Wannier function. The hoppings t and r are the nearest neighbors hoppings of the minimal model. c) Bulk bands of monolayer model. We can observe that there are two $\Gamma_1^+$ irreps in the valence bands, which produce the non-trivial $\mathbb{Z}_2=1$ topological invariant.}
    \label{fig:fig3}
\end{figure*}

\section{$\text{YAsS}$}\label{sec:YAsS}

In the following sections we will explain the electronic structure of each compound in depth. We start our study with YAsS. SG $Pnma$, common to all studied compounds, is formed by 2 fold screw axes $\{C_{2x}|\frac{1}{2}\frac{1}{2}\frac{1}{2}\}$, $\{C_{2y}|0\frac{1}{2}0\}$ and $\{C_{2z}|\frac{1}{2}0\frac{1}{2}\}$, glides $\{m_{x}|\frac{1}{2}\frac{1}{2}\frac{1}{2}\}$ (g${}_{100}$) and $\{m_{z}|\frac{1}{2}0\frac{1}{2}\}$, mirror $\{m_{y}|0\frac{1}{2}0\}$ and inversion symmetry at the origin $\{I|000\}$. Since they are non-magnetic, time-reversal symmetry (TRS) is also present. The crystal structure and relevant glide symmetry are shown in Fig.~\ref{fig:fig1}a.  In Fig.~\ref{fig:fig1}b, we display the Brillouin zone (BZ) of space group $Pnma$, indicating the high-symmetry points and the location of the glide symmetry, which leaves the $k_x = 0, \pi$ planes invariant. In Fig.~\ref{fig:fig1}c we show the bulk electronic band dispersion. In blue continuous lines we show the calculation in the absence of SOC (NSOC) and in black dashed lines we show the calculation including SOC. The two band structures are relatively close to each other, with one difference: the NSOC calculation shows band crossings while the SOC calculation is gapped (see inset  Fig~\ref{fig:fig1}c). In order to characterize the topological properties of the system, we compute the symmetry indicators (SIs) both in the presence and absence of SOC by using the Fu-Kane-like formula for the $\mathbb{Z}_4$ indicator \cite{Fu-Kane,Zhida_NSOC,SongSI}:
\begin{table}
    \centering
    \begin{tabular}{c|c|c}
        $\boldsymbol{K}$ & NSOC & SOC \\\hline
        $\Gamma$ & 4\gup$\oplus$1\gum$\oplus$2\gdp$\oplus$3\gdm$\oplus$
                   2\gdp$\oplus$3\gdm$\oplus$4\gcp$\oplus$1\gcm & 12\gsp$\oplus$8\gsm \\
        $U$ & 4\yup$\oplus$2\yum$\oplus$1\ydp$\oplus$3\ydm & 5\ysp$\oplus$5\ysm \\
    \end{tabular}
    \caption{Irreducible representations in the absence and presence of SOC at $\Gamma$ and $U$ k-points. The band inversion happens at $\Gamma$, where there is an imbalance of even/odd eigenvalues in the irrep decomposition. 
    }
    \label{tab:irreps}
\end{table}

\begin{equation}
    \mathbb{Z}_{4} =\sum_{K\in \text{TRIM}}\frac{n^-_K-n^+_K}{2}\text{ mod 4},
\end{equation}
where $n^-_K$ ($n^+_K$) denote the multiplicity of irreducible representations (irreps) with -1 (+1) eigenvalue of inversion symmetry. We show in Table~\ref{tab:irreps} the irreps of the last disconnected set of bands, up to the Fermi level, both with and without SOC. We only list the irreps at $\Gamma$ and $U$, since they are the only high-symmetry k-points that can host a band inversion. We compute the symmetry indicator and get that $\mathbb{Z}_4=2$, both with and without SOC. In both cases, the non-trivial symmetry indicator stems from a band inversion at the $\Gamma$ point, with 4 more states with inversion symmetry eigenvalue +1 than states with eigenvalue -1. Interestingly, while the SOC gaps the crossings of the NSOC case, it does not alter the symmetry indicator. This is different from usual topological insulators, where the band inversion is driven by SOC \cite{Moore2010}. This suggests that we can construct a minimal NSOC model for the system that describes the main topological properties.

\begin{figure*}
    \centering
    \includegraphics[width=\linewidth]{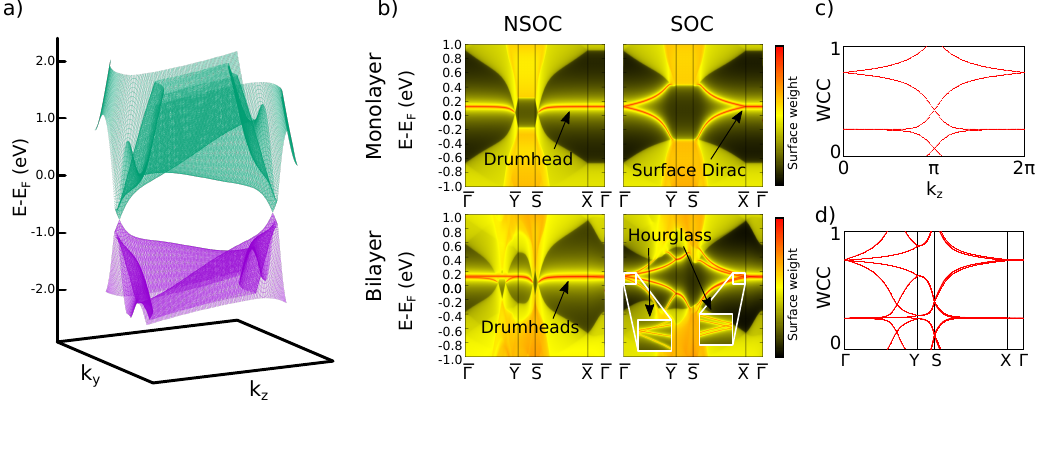}
    \caption{Bulk and surface properties of the monolayer and bilayer models. a) Bulk band dispersion of last valence and first conduction bands of the monolayer model, showing the two Dirac crossings. b) Surface spectrum of the models. In the top row, we show the monolayer surface spectrum without and with SOC. In the bottom row, we show the bilayer surface spectrum, without and with SOC. For illustrative purposes, we increased the hoppings by a factor of 100 in the plots. 
    c) Wilson loops of the monolayer (top) and bilayer (bottom) models. }
    \label{fig:fig4}
\end{figure*}

In the absence of SOC, it was shown that the SI $\mathbb{Z}_4=2$ corresponds to topological invariant $z_2'=\frac{\mathbb{Z}_4}{2}=1$ \cite{Zhida_NSOC}, which implies the presence of an odd number of topologically charged nodal lines in each half of the BZ (see Fig.~\ref{fig:fig1}b). In our case, we find 1 nodal at each side of the $k_z=\pi$ plane (see the $k_y$ projection in Fig.~\ref{fig:fig1}b).
The topological charge of these nodal lines is of $\mathbb{Z}_2$ type, and it can be understood as the Berry phase along a loop that is `chained' to the nodal line \cite{Fang_2016}. In Fig.~\ref{fig:fig1}b we show the path in momentum space that we used to determine the topological charge of one of the nodal lines (long dashed red line), with the resulting Berry phase of $\pi$. We similarly checked the other nodal line and we obtained the same result. To further confirm the $\mathbb{Z}_2$ nature of the charge, we computed the Berry phase on a path that chains both nodal lines together (short dashed blue line) and obtained a Berry phase of 0, which is the result of the sum of the two $z'_2=1$ topological invariants of both nodal lines. These nodal lines are expected to present surface states stemming out of their projection in the surface, similar to Fermi arcs stemming out of Weyl nodes, called drumhead states \cite{Fang_nodal,Weyl,CoS2_theory}. We can see the drumhead states of YAsS in Fig.~\ref{fig:fig1}d left, in the (0-11) surface.

In the presence of SOC, the $\mathbb{Z}_4=2$ SI implies the presence of glide and TRS protected surface states known as hourglass fermions \cite{Hourglass}, which are topological surface states whose characteristic 'hourglass' shape is symmetry enforced.
In Fig.~\ref{fig:fig1}d we show the surface spectrum of YAsS, with the characteristic hourglass fermions. In Appendix~\ref{app:app1} we give further details on the origin of hourglass fermions and their symmetry protection.

\subsection{Minimal model}\label{sec:sec4}

To clarify the emergence of the topological surface states and their robustness against electronic correlations, we construct a minimal model from the orbitals driving the band topology. Within this framework, the hourglass dispersion decomposes into an abstract layer construction consisting of two $\mathbb{Z}_2=1$ two‑dimensional topological insulators—one centered at the origin and the other at the unit‑cell midpoint—coupled by a glide symmetry \cite{SongSI}. In particular, one can observe in Fig.~\ref{fig:fig1}a that As atoms are arranged in two layers—one at the origin and the other at the center of the unit cell—and in Fig.~\ref{fig:fig3}a, the NSOC band structure of YAsS shows that the dominant weight near the Fermi level originates from the As $p_z$ and $p_y$ orbitals, as confirmed by the spatial distributions of the corresponding Wannier functions in Fig.~\ref{fig:fig3}b. This implies that the minimal model can be constructed from As $p_z$ and $p_y$ orbitals located at those atomic positions.
This model will serve as a basis for explaining the basic topological properties of the compunds in the family that share the same orbital content close to the Fermi level. As we will see, this is the case for SmAsS, but not for LaAsS. For SmAsS, this construction can be extended to include the f‑electron degrees of freedom stemming from Sm $4f$ orbitals.
In what follows, we explicitly build this model first for the As monolayer and then for the As bilayer, which together form the full model.

\subsection{As monolayer $\mathbb{Z}_2=1$}

We construct maximally localized Wannier functions solely for As $p_z$ and $p_y$ orbitals from the NSOC DFT calculation, and we keep only first nearest neighbors of the interpolated Hamiltonian. This model is depicted in Fig.~\ref{fig:fig3}b, the bulk band structure is shown in Fig.~\ref{fig:fig3}c and full details on Wyckoff position and numerical values of the hoppings can be found in Appendix~\ref{app:app2}.
Using the Mathematica package MagneticTB \cite{Mathematica,MagneticTB}, we compute the inversion eigenvalues of the monolayer model at half filling, confirming a $\mathbb{Z}_2=1$ topological invariant. Fig.~\ref{fig:fig3}c presents the irreducible representations that drive the band inversion, reproducing the minimal inversion pattern of the full system. By combining this invariant with the absence of SOC, the system is forced into a semimetallic phase, exhibiting two Dirac crossings at $(k_y,k_z)=\pm(0, 0.753)\text{\AA}$ (see Fig.~\ref{fig:fig4}a). Since the 2D Dirac points can be viewed as flat nodal lines along $k_x$, their surface states are analogous to the drumhead states that emerge from projecting 3D nodal lines. \cite{Fang_2016}.

We incorporate spin–orbit coupling via the dominant Wannier term, $H_{\text{SOC}}=\lambda i\sigma_y$, acting on-site between the $p_y$ and $p_z$ orbitals. To open a substantial gap and clearly resolve the surface states, we set $\lambda = 0.5,\mathrm{eV}$. This term gaps the Dirac nodes while leaving the occupied irreducible representations unchanged, so the $\mathbb{Z}_2$ invariant is preserved.
To further validate this result, we compute the Wilson loops by integrating along $k_y$ and plotting versus $k_z$ (see Fig.~\ref{fig:fig4}c), which exhibit the characteristic winding of a $\mathbb{Z}_2=1$ 2D TI. We also calculate the surface spectrum, revealing the expected surface Dirac nodes (see Fig.~\ref{fig:fig4}b, top‑right panel).

\subsection{As bilayer: $\mathbb{Z}_4=2$}\label{subsec:full_bilayer_model}

Starting from the monolayer tight‑binding Hamiltonian, we build the full (bilayer) model by appending a second monolayer related via the glide symmetry $g_{100}$. All intralayer hopping amplitudes are left unchanged, while the interlayer couplings are then obtained by extracting nearest‑neighbor hopping parameters from the Wannierization. In the absence of SOC, surface states originating from both nodal lines emerge (Fig.~\ref{fig:fig4}b, bottom left). Upon introducing SOC, a gap opens and these surface states split, yielding the characteristic hourglass‑fermion dispersion (Fig.~\ref{fig:fig4}b, bottom right). To verify this, we compute the Wilson loop spectrum integrated along $k_x$ following the same surface momentum path as in Fig.~\ref{fig:fig4}b; the resulting bands (Fig.~\ref{fig:fig4}d) exhibit the expected winding for a $\mathbb{Z}_4=2$ topological invariant \cite{Hourglass}.

In summary, this minimal model faithfully reproduces all of YAsS’s topological features at the Fermi level. Furthermore, because SmAsS exhibits a comparable orbital character at Fermi level, the same model can serve as an effective proxy for other members of the family with the same orbital content.

\section{$\text{SmAsS}$}\label{sec:SmAsS}

\begin{figure*}
    \centering
    \includegraphics[width=\linewidth]{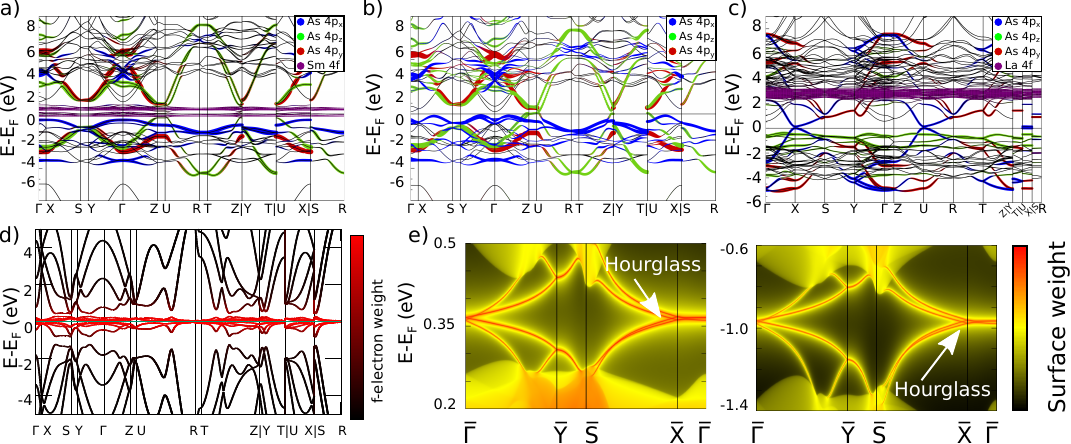}
    \caption{Bulk and surface spectrum including the $f$-electrons. a) Bulk bands of SmAsS. In color, the orbital weight of As 4$p_i$ orbitals  ($i=x, y, z$) and Sm $4f$ orbitals. b) Bulk bands of SmAsS with the Sm $4f-$electrons in the core. We can see how the orbital weight of As $4p$ orbitals is identical. c) Bulk band structure of LaAsS. Notice the difference in the orbital weight close to the Fermi level compared to Y/SmAsS. d) Bulk band structure of the heavy-fermion model. The colorbar represents the weight of the f-orbitals. e) Surface spectrum of the heavy-fermion model in both topological gaps.}
    \label{fig:fig5}
\end{figure*}

To explore the interplay between topology and $f$‑electron–driven heavy‑fermion physics, we extend the minimal model developed in previous section to SmAsS.
We carried out density functional theory (DFT) calculations using the meta–Becke–Johnson potential, which has been demonstrated to reliably reproduce electronic band gaps \cite{MBJ}. The resulting bulk band structure is shown in Fig.~\ref{fig:fig5}a, with the Sm $4f$ orbital character highlighted in purple at the Fermi level. Experimentally, we measure a large effective electron mass (see Fig.~\ref{fig:exp}d), in agreement with the flat‑band dispersion predicted by our DFT results.
The As $4p$ orbital weights remain essentially unchanged from those in YAsS (see Fig.~\ref{fig:fig3}a and Fig.~\ref{fig:fig5}a). Moreover, we perform DFT calculations of SmAsS setting the Sm $4f$-electrons in the core (see Fig.~\ref{fig:fig5}b) and we find the orbital weight of As $4p$ orbitals remains unchanged too. Consequently, introducing $f$‑electrons does not affect the band inversion at the $\Gamma$ point that gives rise to the nontrivial topology. Instead, $f$‑electron occupancy shifts the topological gap downward by roughly 0.4 eV—an effect we verify through calculated SIs. A comparable topological phase also appears about 1 eV above the Fermi level \cite{PhysRevB.101.075138,lai2018weyl}. Because this downward shift brings the topological gap into closer proximity to the Fermi energy, it becomes more accessible to spectroscopic techniques such as angle-resolved photoemission spectroscopy (ARPES), making SmAsS an ideal platform for probing the interplay between band topology and %correlated electrons.
strong electron correlation.

\subsection{Anderson lattice model}

Since the orbital weight of As $4p$ orbitals in SmAsS, both taking into account explicitly the $f$-electrons and placing them on the core, is very similar to that of YAsS, we can extend the model derived previously to account for the $f$-electrons. This will allow us to probe the robustness of the topology and validate our DFT results. Specifically, we employ an Anderson lattice Hamiltonian with itinerant As‑derived electrons and strongly correlated Sm $f$‑electrons respecting the symmetries of space group $Pnma$ (No. 62). We stress that space group symmetry constraints play a central role in our treatment of the strong correlation effects.

The Hamiltonian of the itinerant electrons is
    $H_c = \sum_k \Psi_k^\dagger \left( h({\mathbf k}) - \mu \right) \Psi_k$
where $\mu$ is chemical potential, $h(\mathbf k)$ is the single-particle minimal model described in Sec.~\ref{sec:sec4}, and $\Psi = (c_{q_1 \uparrow}, c_{q_1 \downarrow},\dots,c_{q_8 \uparrow}, c_{q_8 \downarrow})$ is the annihilation operator of itinerant electrons at $\rm As$ sublattice sites with spin degrees of freedom.

We assume the $f$ electrons in the $\rm Sm$ atoms are at the energy level $\epsilon_f$ and consider a Hubbard interaction in the infinite $U$ limit~\cite{Hewson1997}.
Here the system is effectively described by an auxiliary-boson representation.
This model is solved at the saddle-point level, which formally develops in a large-$N$ limit ~\cite{Hewson1997, lai2018weyl, chen2022topological}. The effective Hamiltonian is given as follows:
\begin{equation}
\label{eqn:model}
\begin{split}
    H=H_c+\sum_{j,\alpha,\sigma} \epsilon_f f_{j\alpha \sigma}^\dagger f_{j\alpha \sigma} +rV \left[c_{j\alpha \sigma}^{\dagger} f_{j\alpha \sigma}+\text { h.c. }\right] \\
    + \lambda \sum_j (n_{f,j} + 8r^2 - Q)    \, . 
\end{split}
\end{equation}
where $j$, $\alpha$ and $\sigma$ stand for unit cell position, sublatttice and the spin degrees of freedom, respective. The factor $8$ before $r^2$ refers to the number of sublattices in each unit cell. 
The $f$ electron particle number $n_{f,j}=\sum_{\alpha \sigma}f^\dagger_{j\alpha \sigma}f_{j\alpha \sigma}$ is %constraint
constrained by the Lagrange multiplier $\lambda$ term and $r=\sqrt{Z}$ is the square-root of the quasi-particle weight. A nonzero value of quasiparticle weight indicates the emergence of the Kondo resonance, which appears in the vicinity of the Fermi energy. We choose $\epsilon_f = -1$, $V = 1$ and $\mu = 0$, $Q=8$ to maintain the half filling constraint on each sites. The orbital-resolved bulk band structures and the corresponding surface spectrum are shown in Fig.~\ref{fig:fig5}d and Fig.~\ref{fig:fig5}e, respectively. 

The $f$-electron level is pinned to the Fermi level due to the half filling constraint. From the hybridization with the $p$-orbital bands, the $f$ bands split and move the topological gap; one above and one below the Fermi level, maintaining the $\mathbb{Z}_4=2$ topological invariant. The model confirms the DFT prediction that the localized $f$-electrons do not destroy the topological gaps, but move them in energy.

\section{$\text{LaAsS}$}\label{sec:LaAsS}

Our experimental results reveal a new crystal structure for LaAsS, distinct from that previously reported \cite{CEOLIN1977137}. While it crystallizes in space group $Pnma$, like Y/SmAsS, we observe a $\sqrt{2}\times\sqrt{2}$ in-plane modulation perpendicular to the long axis. Fig.~\ref{fig:exp}c compares the structures of LaAsS and SmAsS, highlighting the distortion and rotation of the As lattice in LaAsS relative to SmAsS.

Fig.~\ref{fig:fig5}c presents the bulk band structure of LaAsS, calculated using the meta–Becke–Johnson potential. Due to the structural distortion, maintaining the conventional $Pnma$ notation requires assigning the long-axis direction (out-of-plane) along $x$ for Y/SmAsS and along $z$ for LaAsS. In both compounds, the out-of-plane orbitals ($p_y$ for Y/SmAsS, $p_z$ for LaAsS) lie away from the Fermi level. However, the in-plane orbital character differs significantly: while both $p_x$ and $p_z$ contribute in Y/SmAsS, only $p_x$ contributes in LaAsS.

Following the analysis for Y/SmAsS, we computed the irreducible representations at $\Gamma$ and $U$ in order to extract the $\mathbb Z_4$ SI. We show the irreps in Table~\ref{tab:irreps_La}. Compared to Y/SmAsS, it presents the same band inversion at the $\Gamma$ point. However, there is an extra band inversion at the $U$ point, which results in a $\mathbb Z_4=0$ SI, thus rendering the system trivial. We repeated the calculation on the un-optimized structure, that is, the experimentally determined structure, which results in a reordering of the irreps at $U$ and a $\mathbb Z_4=2$ SI, as Y/SmAsS. Since the gap of LaAsS is very small, we expect it to be very sensitive both to modeling (DFT calculations with different exchange-correlation approximations) and experimental conditions (external pressure, strain or growing conditions).

\begin{table}
    \centering
    \begin{tabular}{c|c}
        $\boldsymbol{K}$ & Irreps \\\hline
        $\Gamma$  & 12\gsp$\oplus$8\gsm \\
        $U$ (optimized)  & 4\ysp$\oplus$6\ysm \\
        $U$ (experimental)  & 5\ysp$\oplus$5\ysm 
    \end{tabular}
    \caption{Irreducible representations of LaAsS at $\Gamma$ and $U$ k-points. The band inversion at $\Gamma$ remains, but there is an extra band inversion at $U$, which renders the system trivial.}
    \label{tab:irreps_La}
\end{table}

\section{Conclusions}\label{sec:conc}

We have identified RAsS (R = Y, La, Sm) as a new family of topological crystalline insulators featuring glide-symmetry-protected hourglass fermions. Our structural analysis clarified their accurate crystal structure and our analysis based on DFT and TQC revealed nontrivial topological invariants in YAsS and SmAsS, with LaAsS near a topological transition. We show that both YAsS and SmAsS present topologically protected surface states in the surface preserving the glide symmetry. SmAsS stands out as a strong candidate for a heavy-fermion topological crystalline insulator, where strong electronic correlations and topology converge. Density functional theory and tight-binding calculations show that their topological properties can be captured with a minimal model. Crucially, in SmAsS, the surface states persist despite $f$-electron interactions and shift downward in energy, enhancing their accessibility for spectroscopic studies, such as ARPES. These results expand the landscape of strongly correlated topological insulating materials beyond the two reported ones (SmB${}_6$ and YB${}_{12}$), 
while unveiling a new mechanism underlying such materials in which symmetry constraints work in tandem with strong correlations, and offer a platform for studying interaction-driven phenomena.

\section{Methods}\label{sec:methods}

\subsection{Calculations}

We performed DFT calculations as implemented in the Vienna ab initio simulation package (VASP) \cite{VASP1, VASP2, VASP3, VASP4} for structural optimization and determination of symmetry indicators. The interaction between the ion cores and valence electrons was treated by the projector augmented-wave method \cite{PAW}, the generalized gradient approximation (GGA) was employed for the exchange-correlation potential with the Perdew–Burke–Ernzerhof for solid parameterization \cite{PBE}, and the spin-orbit coupling (SOC) was considered based on the second variation method \cite{DFT-SOC}. A $\Gamma$-centered Monkhorst-Pack k-point grid of ($5 \times 9 \times 9$) was used for reciprocal space integration for Y/SmAsS and ($9 \times 9 \times 5$) for LaAsS, and 500 eV energy cutoff of the plane-wave expansion. We ensured convergence up to $10^{-5}$eV per unit cell.
We performed geometric optimization of the structure until the forces in the atoms were smaller than $10^{-2}$eV/\AA.
We computed the irreducible representations of the occupied set of bands in all systems using Vasptotrace software and computed the SIs as implemented in the Bilbao Crystallographic Server \cite{BCS1,BCS2,BCS3}. To enforce symmetry constraints, we constructed maximally localized Wannier functions from a DFT calculation as implemented in FPLO \cite{FPLO}, and we run surface state calculations following the iterative Green's function method as implemented in WannierTools \cite{WannierTools}.

Eq.~(\ref{eqn:model}) is solved at the saddle point level, where the self-consistent equations are derived by taking the partial derivatives of the effective Hamiltonian with respect to $\lambda$ and $r$, respectively. It leads to:
\begin{align}
    \langle n_{f,j}\rangle + 8r^2 &= Q \\
    V \sum_{\alpha,\sigma}\langle c_{j\alpha \sigma}^{\dagger} f_{j\alpha \sigma}\rangle &= -16\lambda r 
\end{align}
where $\langle\cdot\rangle$ means averaging over all sites in the unit cell. These equations are solved on a $40\times 40 \times 10$ grid. We find $r = 0.805$, $\lambda = 0.503$ for the parameters we adopted.
\\

\subsection{Synthesis and characterization}

Glassy carbon crucibles were heat-treated under a dynamic vacuum before use. The starting materials were: yttrium pieces (ChemPur, 99.9\%), lantanum chunks (Alfa Aesar, 99.9\%), samarium powder (ChemPur, 99.9\%), arsenic lump (Alfa Aesar, Puratronic, 99.9999\%), sulfur pieces (Alfa Aesar, 99.999\%). Well-ground LiCl (Thermo Scientific, Ultra Dry, 99.9\%) and RbCl (Thermo Scientific, 99.8\%) with a ratio of 1:1 were used as a flux. Then, all starting materials with the total mass of 1g for RE, As, and S (ratio 1:1:1), and 2.5g of salt mixture (ratio 1:1) were loaded in a glassy carbon, covered by flux material, and enclosed in a Ta ampule using the arc-welding machine. Prior to synthesis, the salts were pre-heated inside the glovebox to remove moisture. Ta ampule with glassy carbon crucible was placed in a vertical furnace and heated with the following temperature gradient – for 12 h to 900 C$^\circ$, stayed for 168 h at this temperature, and cooled down to room temperature for 240 h. Residual salt was washed away with deionized water. Preparation and synthesis took place in a glove box (O$_2$, H$_2$O $\leq$ 0.1 ppm).

The lattice parameters were established by using high resolution powder diffraction data ($\lambda$ = 0.35466 \AA) recorded at ID22 beamline at European Synchrotron Radiation Facility (ESRF). For single crystal experiments, small pieces were used ($\sim 20~\mu$m). The single crystal diffraction data were collected using a Rigaku AFC7 diffractometer (equipped with a Saturn 724+ CCD detector) and a Bruker Apex II diffractometer, both with MoK$_\alpha$ radiation ($\lambda$ = 0.71073 \AA). For YAsS and SmAsS, single crystal diffraction data were collected using  with Mo  The complete crystallographic information is given in Supplementary Tables \ref{T1}, \ref{T2}, and \ref{T3}.

The single crystals of YAsS, LaAsS and SmAsS were additionally analyzed by energy-dispersive X-ray spectroscopy with a Jeol JSM 6610 scanning electron microscope equipped with an UltraDry EDS detector (ThermoFisher NSS7). The semi-quantitative analysis was performed with 30 keV acceleration voltage. No impurity elements were observed, confirming that no reaction with the crucible took place during synthesis. The experimentally determined element ratios were in a good agreement with the 1:1:1 stoichiometry. 

Temperature- and field-dependent magnetic measurements were conducted in a Quantum Design (QD) Magnetic Properties Measurement System (see main text Fig.~\ref{fig:exp}). Several single crystals of YAsS and SmAsS were mounted on a quartz capillary. Magnetic moment was measured at temperatures ranging from 2 to 600 K and in magnetic fields up to $H = 7$ T. The specific heat data were collected on a QD Physical Property Measurement System (PPMS) from $T = 0.4$ K to $T = 100$ K in $H = 0$ and $H = 9$ T magnetic fields. For measurements of electrical resistivity,  micro-scale devices were fabricated out of a YAsS, SmAsS or LaAsS single crystal by using a plasma focused-ion-beam (FIB). AC electrical resistivity measurements were performed by a QD PPMS, using a standard four-probe technique at temperatures between $T = 2$ and 300 K in $H = 0$ and $H = 9$ T applied magnetic field. A current pulse of 0.01 mA with frequency 93 Hz for 1 s was applied.

\section{Acknowledgments}

M.G.V. thanks support to PID2022-142008NB-I00 project funded by  MICIU/AEI/10.13039/501100011033 and FEDER, UE, Canada Excellence Research Chairs Program for Topological Quantum Matter, NSERC Quantum Alliance France-Canada and to Diputación Foral de Gipuzkoa Programa Mujeres y Ciencia. This work was supported by the Deutsche Forschungsgemeinschaft (DFG) through QUAST-FOR5249 and the Würzburg-Dresden Cluster of Excellence on Complexity and Topology in Quantum Matter, ct.qmat (EXC 2147, Project ID 390858490). We also acknowledge funding from the EU NextGenerationEU/PRTR-C17.I1, as well as by the IKUR Strategy under the collaboration agreement between Ikerbasque Foundation and DIPC on behalf of the Department of Education of the Basque Government. E.S. is grateful for the support of the Christiane N\"usslein-Volhard-Stiftung E.S., N. Z. and M. K. acknowledge the support of the Boehringer Ingelheim Plus 3 Program. N.Z. is grateful for the support of the Humboldt Foundation through the Philipp Schwartz Initiative.
Work at Rice has primarily been supported by the Air Force Office of Scientific Research under Grant No. FA9550-21-1-0356 (correlated electron model construction, Y.F. and L.C.), by the National Science Foundation under Grant No. DMR-2220603 (model calculations, Y.F. and L.C.),
and by the Robert A. Welch Foundation Grant No. C-1411 and the Vannevar Bush Faculty Fellowship ONR-VB N00014-23-1-2870 (Q.S.). M. G. V. and Q.S. acknowledge the hospitality of the Kavli Institute for Theoretical Physics, UCSB, supported in part
by the National Science Foundation under Grant No. NSF PHY-1748958, during the program ``A Quantum Universe in
a Crystal: Symmetry and Topology across the Correlation Spectrum."  

\bibliographystyle{unsrt}
\bibliography{biblio}

\onecolumngrid

\clearpage
\appendix

\setcounter{figure}{0}
\setcounter{table}{0}
\renewcommand{\figurename}{FIG.}
\renewcommand{\tablename}{TABLE S.}
\renewcommand{\thefigure}{S\arabic{figure}}

\section{Implications of $\mathbb{Z}_4=2$ symmetry indicator and surface spectrum}\label{app:app1}

\begin{figure}[b!]
    \centering
    \includegraphics[width=0.7\linewidth]{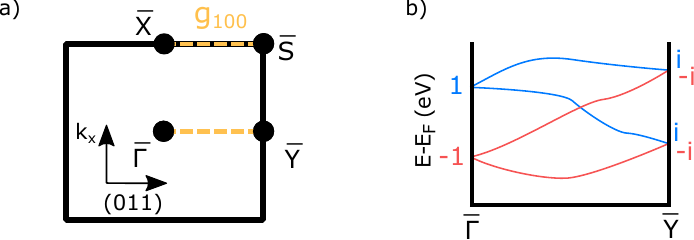}
    \caption{Surface properties of a glide-symmetric surface. a) Surface BZ highlighting glide symmetry protected paths. b) Schematic origin for the hourglass fermion dispersion in the surface spectrum.}
    \label{fig:si_fig1}
\end{figure}

In this appendix, we illustrate the symmetry protection of hourglass fermions in SG $Pnma$. The glide symmetry g${}_{100}$ leaves the planes $k_x^p=0,\pi$ invariant. We can get the eigenvalues of this symmetry at a general k-point $(k_x^p,k_y,k_z)$ by noticing that:
\begin{equation}
    \left(\{m_{x}|\frac{1}{2}\frac{1}{2}\frac{1}{2}\}\right)^2=\{E|011\}=R_y+R_z,
\end{equation}
where $R_y+R_z$ is an integer lattice translation in the $yz$ direction. Restricting to the $k_z=0$ plane, the eigenvalues of the glide are $\pm\sqrt{e^{ik_y}}=\pm e^{ik_y/2}$. This is also true for the states in the $(01\bar 1)$ surface BZ in the $\bar{\Gamma}-\bar{Y}=(0,\bar{k}_y)$ and $\bar{X}-\bar{S}=(\pi,\bar{k}_y)$ paths. Focusing on the first path, we compute the eigenvalues of surface bands at both extremes. At $\bar{\Gamma}=(0,0)$, the eigenvalues are $\pm e^{i0/2}=\pm 1$. Due to TRS-enforced Kramers degeneracy, the bands at $\bar{\Gamma}$ are doubly degenerate, both of which have to share either $+1$ or $-1$ glide eigenvalue. At the other extreme, at $\bar{Y}=(0,\pi)$, the eigenvalues of the glide symmetry are $\pm e^{i\pi/2}=\pm i$. In this case, TRS maps the state with eigenvalue $+i$ to the state with eigenvalue $-i$, thus, Kramers pairs need to have opposite glide eigenvalues. Since the glide is a symmetry of the whole line, we can follow the bands from $\bar{\Gamma}$ to $\bar{Y}$ by labeling the eigenvalues. Once away from the $\bar{\Gamma}$ point, TRS is not a symmetry of the little group and thus Kramers pairs will split. In particular, the states that emerge from eigenvalue $+1$ ($-1$) will evolve to eigenvalue $+e^{ik_y/2}$ ($-e^{ik_y/2}$). At the end of the path, at $\bar{Y}$, the states with eigenvalue $+e^{ik_y/2}$ ($-e^{ik_y/2}$) will have eigenvalue $+i$ ($-i$). Since TRS forces Kramers pairs to have opposite eigenvalues, the Kramers partners must change along the path, i.e., the bands \emph{cross} along the path. Notice that bands that must cross carry different eigenvalues in the path, so their crossing is glide-symmetry protected. This procedure holds for the $\bar{X}-\bar{S}$ path as well. For the other two high-symmetry paths, i.e. $\bar{\Gamma}-\bar{X}$ and $\bar{Y}-\bar{S}$, the little group of the line contains the combination of the glide and TRS, so that the bands are Kramers-degenerate along the whole line. A sketch of the discussion is shown in Fig.~\ref{fig:si_fig1}a and Fig.~\ref{fig:si_fig1}b.

 \begin{table}[h]
    \centering
    \begin{tabular}{c|ccc}
         & $p_z$-$p_z$ (eV)  & $p_z$-$p_y$ (eV)  & $p_y$-$p_y$ (eV) \\ \hline
        on-site & 0.0 & - & 0.264 \\ \hline
        t & 0.851 & 1.949 & 1.325 \\ \hline
        r & 0.981 & 1.368 & 0.671 \\ \hline
    \end{tabular}
    \caption{Numerical values of the minimal monolayer TB model. All terms are written in units of eV.}
    \label{tab:mono_hoppings}
\end{table}

\begin{table}
\renewcommand{\tablename}{TABLE S.}
    \centering
    \begin{tabular}{c|c|c|c|c}
        $\boldsymbol{K}$ & $\Gamma$ & $Y$ & $T$  & $Z$ \\\hline
        Inversion Eigenvalues & 2+  &  1+ 1-  & 1+ 1-  & 1+ 1- \\
    \end{tabular}
    \caption{Inversion eigenvalues of the minimal model at high-symmetry k-points. The band inversion happens at $\Gamma$, where there is an imbalance of even/odd eigenvalues in the irrep decomposition, just as in the DFT result in Table~\ref{tab:irreps}.}
    \label{tab:irreps_minimal}
\end{table}

\begin{table}
\renewcommand{\tablename}{TABLE S.}
    \centering
    \begin{tabular}{c|cccc}
         & $p_z$-$p_z$ (meV)  & $p_z$-$p_y$ (meV)  & $p_y$-$p_z$ (meV) & $p_y$-$p_y$ (meV) \\ \hline
        v & 1.013 & 0.248 & 0.440 & 0.949 \\
    \end{tabular}
    \caption{Numerical values of the interlayer coupling. All terms are written in units of meV.}
    \label{tab:interlayer_hoppings}
\end{table}

\section{Minimal model details}\label{app:app2}

In this section we provide full details on the minimal model. First, the only symmetries preserving the monolayer are inversion symmetry and $\{m_{y}|0\frac{1}{2}0\}$, resulting in a monolayer space group of $P2_1/m$ (\#11). The Wyckoff positions for the monolayer model are $q_0 = (0.499, 0.750, 0.217)$ and $q_2 = -q_0$. Each As atom has two neighbors, with hopping amplitudes denoted by $t$ and $r$, ordered by increasing distance (see main text Fig.~\ref{fig:fig3}). Each bond involves three types of hopping: $p_z$-$p_z$, $p_z$-$p_y$, and $p_y$-$p_y$. The numerical values of these hopping amplitudes, obtained from Wannierization, are presented in Table~\ref{tab:mono_hoppings}.

Using the Mathematica package MagneticTB \cite{Mathematica,MagneticTB} we compute the inversion eigenvalues of the monolayer model at half filling and display them in Table~\ref{tab:irreps_minimal}, which give rise to a topological invariant $\mathbb{Z}_{2} =1$ as we explain in the main text. In order to couple the layers, we extract the hopping parameters from the wannierization. We only consider the nearest neighbors, which connects $q_0-q_1$ and $q_2-q_3$. They are related by symmetry, thus we only list the hoppings between $q_0-q_1$ in Table~\ref{tab:interlayer_hoppings}.

\section{Crystallographic information}\label{app:app3}

\begin{figure*}
    \centering
    \includegraphics[width=\linewidth]{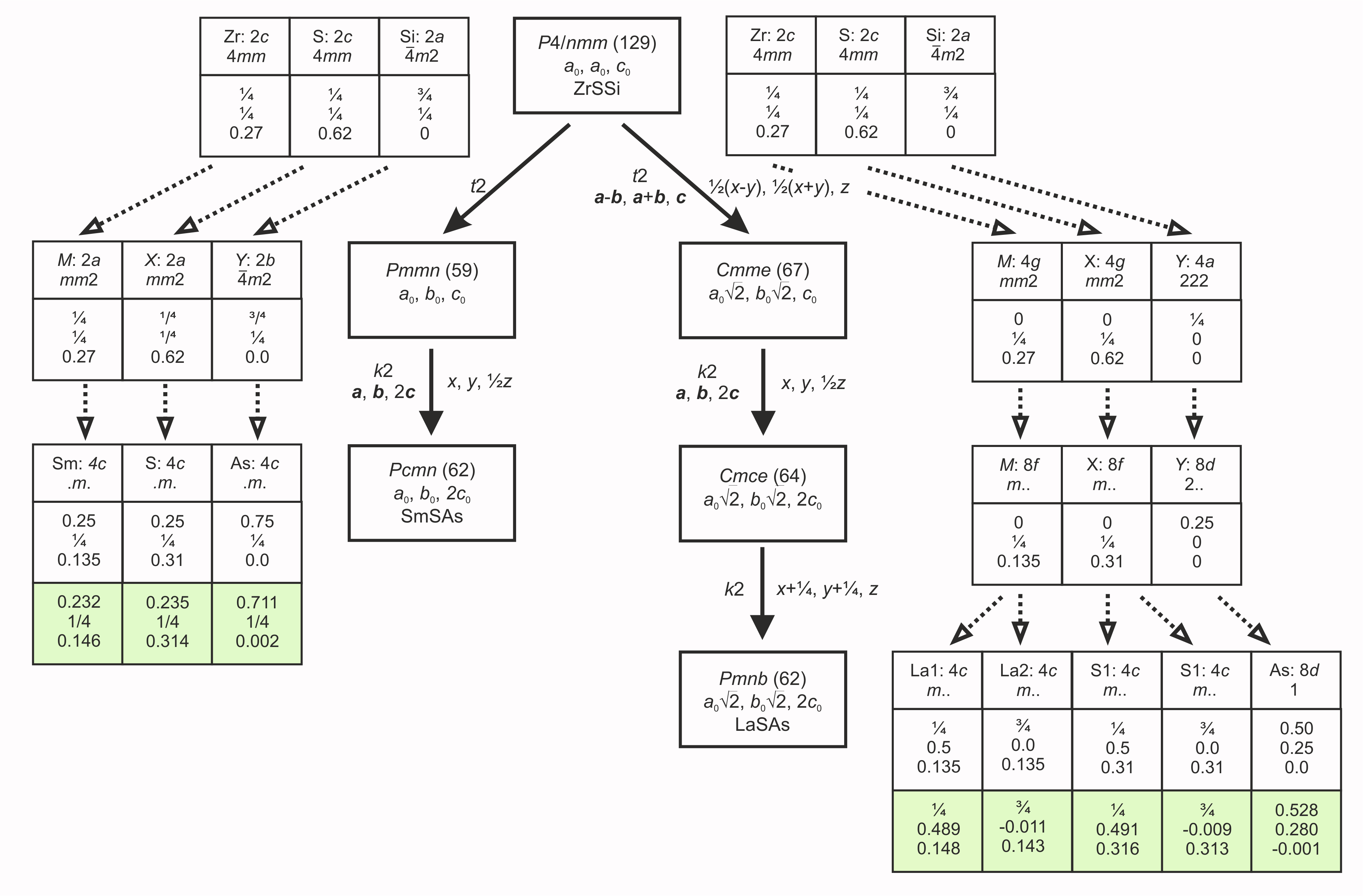}
    \caption{The Bärnighausen trees for LaAsS and SmAsS, showing their relation to the ZrSiS structure type. Note that since it is not possible to have the tree in all-standard setting -- the viewing directions in the tetragonal and orthorhombic crystal system are different. The calculations haven been done using the standard Pnma setting of the space group types.}
    \label{fig:BT}
\end{figure*}

\begin{footnotesize}
\begin{table}
\renewcommand{\tablename}{TABLE S.}
\renewcommand{\arraystretch}{1.1} 
\caption{Crystallographic data of YAsS, SmAsS, and LaAsS (standard setting).}
\begin{tabular}{ l l l l } \hline \hline
Composition	& YAsS	& SmAsS	& LaAsS \\ \hline \hline
Structure type	& SmAsS	& SmAsS	& GdPS \\
Space group	& \multicolumn{3}{c}{$Pnma$} \\
Pearson symbol	& $oP12$	& $oP12$	& $oP24$ \\
Formula units per unit cell, Z & 4	& 4	& 8 \\
Unit cell parameters\footnote{The lattice parameters were established by using high resolved powder diffraction data ($\lambda$ = 0.35466 \AA) recorded at ID22 beamline at European Synchrotron Radiation Facility (ESRF).}	&&& \\
$a$, \AA & 16.88916(6) & 17.1455(2) & 5.7078(6) \\
$b$, \AA & 3.80572(1) & 3.86008(3) & 5.6924(5) \\
$c$, \AA & 3.84082(1) & 3.91298(3) & 17.553(2) \\
Unit cell volume, \AA $^3$ & 246.870(1)	& 258.973(4) & 570.32(8) \\
Calculated density, g/cm$^{-3}$ & 5.287 & 6.600 & 5.728 \\
Diffraction system	& Bruker D8 & Bruker D8 & RIGAKU AFC7 \\
Detector & Apex II CCD & Apex II CCD & Saturn 724+ CCD \\
Radiation $\lambda$ \AA & \multicolumn{3}{c}{MoK$\alpha$, 0.71073} \\
Scan; step / degree; N(images) & 0.3, 4934 & 0.3, 5992 &$\phi$, 0.5, 720 \\
Maximal 2$\theta$ / degree  & 80.35 & 80.30 & 66.23 \\
Measured range in hkl & -30 $\leq$ h $\leq$ 30 & -31 $\leq$ h $\leq$ 31 & -3 $\leq$ h $\leq$ 8 \\
	& -6 $\leq$ k $\leq$ 6 & -7 $\leq$ k $\leq$ 6 & -8 $\leq$ k $\leq$ 8 \\
	& -6 $\leq$ l $\leq$ 6 & -7 $\leq$ l $\leq$ 7 & -26 $\leq$ l $\leq$ 24 \\
Absorption correction & \multicolumn {3}{c}{multi-scan} \\
T(max)/T(min) & 2.13 & 2.20 & 3.95 \\
Absorption coefficient, mm$^{-1}$ & 37.35 & 35.76 & 26.87 \\
N(hkl) measured & 15720 & 14949 & 4727 \\
N(hkl) unique & 921 & 952 & 1224 \\
R(int) & 0.0566 & 0.0438 & 0.0162 \\
N(hkl) observed & 885 & 935 & 1143 \\
Observation criteria & \multicolumn{3}{c}{F(hkl) $\leq$ 4$\sigma$ [F(hkl)]} \\
Twinning law & -1 0 0  & -1 0 0  & 0 1 0  \\
&  0 0 1  &  0 0 1 &  1 0 0 \\
&  0 1 0 & 0 1 0 & 0 0 -1 \\
Twin components ratio & 0.518(2):0.482 & 0.688(3):0.312 & 0.929(4):0.071 \\
Number of refined parameters & 22 & 22 & 35 \\
R1 & 0.0240 & 0.0282 & 0.0339 \\
wR2 & 0.0652 & 0.0722 & 0.0969 \\
Residual peaks (e$^{-}$ \AA$^{-3}$) & 2.04/-2.47 & 2.53/-4.72 & 2.53/-1.98 \\ \hline \hline
\end{tabular}
\label{T1}
\end{table}

\begin{table}
\renewcommand{\tablename}{TABLE S.}
\renewcommand{\arraystretch}{1.3} 
\caption{Atomic coordinates and equivalent displacement parameters (in \AA$^2$) in the YAsS, SmAsS and LaAsS (standard setting).}
\begin{tabular}{ l l l c l l } \hline \hline
Atom & Wyckoff site & $x/a$ & $y/b$ & $z/c$ & Ueq. \\ \hline \hline 			
\multicolumn {6}{c}{YAsS ($Pnma$, SmAsS structure type)} \\
Y & $4c$ & 0.14574(2) & \sfrac{1}{4} & 0.23305(9) & 0.00391(10) \\
As & $4c$ & 0.00146(3) & \sfrac{1}{4} & 0.71659(11) & 0.00427(10) \\
S & $4c$ & 0.31334(5) & \sfrac{1}{4} & 0.2357(2) & 0.00395(14) \\
\multicolumn {6}{c}{SmAsS ($Pnma$, SmAsS structure type)} \\
Sm & $4c$ & 0.14572(2) & \sfrac{1}{4} & 0.23234(8) & 0.00481(9) \\
As & $4c$ & 0.00172(5) & \sfrac{1}{4} & 0.71092(19) & 0.00576(12) \\
S & $4c$ & 0.31399(8) & \sfrac{1}{4} & 0.2344(4) & 0.0054(2) \\
\multicolumn {6}{c}{LaAsS ($Pnma$, GdPS structure type)} \\
La1 & $4c$ & 0.01102(7) & \sfrac{1}{4} & 0.35162(2) & 0.00856(14) \\
La2 & $4c$ & 0.48941(7) & \sfrac{1}{4} & 0.64314(2) & 0.00897(14) \\
As & $8d$ & 0.28045(11) & 0.02824(10) & 0.00131(2) & 0.01224(15) \\
S1 & $4c$ & 0.0089(3) & \sfrac{1}{4} & 0.18347(11) & 0.0089(3) \\
S2 & $4c$ & 0.4908(3) & \sfrac{1}{4} & 0.81282(11) & 0.0092(3) \\ \hline

\end{tabular}
\label{T2}
\end{table}

\begin{table}
\renewcommand{\tablename}{TABLE S.}
\renewcommand{\arraystretch}{1.3} 
\caption{Anisotropic displacement parameters  (in \AA$^2$) for YAsS, SmAsS and LaAsS (standard setting).}
\begin{tabular}{ l l l l c l c } \hline \hline
 
Atom & $U_{11}$ & $U_{22}$ & $U_{33}$ & $U_{23}$ & $U_{13}$ & $U_{12}$ \\ \hline \hline
\multicolumn {6}{c}{YAsS ($Pnma$, SmAsS structure type)} \\
Y & 0.00198(12) & 0.0083(3) & 0.00148(18) & 0 & 0.00008(8) & 0 \\
As & 0.00155(14) & 0.0090(2) & 0.00231(16) & 0 & -0.00025(10) & 0 \\
S & 0.0015(3) & 0.0085(6) & 0.0018(5) & 0 & -0.00018(19) & 0 \\
\multicolumn {6}{c}{SmAsS ($Pnma$, SmAsS structure type)} \\
Sm & 0.00292(12) & 0.00524(19) & 0.00628(18) & 0 & 0.00014(7) & 0 \\
As & 0.0029(2) & 0.0063(3) & 0.0081(3) & 0 & 0.0001(2) & 0 \\
S & 0.0033(4) & 0.0058(7) & 0.0070(7) & 0 & 0.0001(3) & 0 \\
\multicolumn {6}{c}{LaAsS ($Pnma$, GdPS structure type)} \\
La1 & 0.0086(2) & 0.0077(2) & 0.0094(2) & 0 & 0.00022(10) & 0 \\
La2 & 0.0088(2) & 0.0076(2) & 0.0105(2) & 0 &-0.00004(10) & 0 \\
As & 0.0136(3) & 0.0127(3) & 0.0105(3) & 0.00106(19) & 0.00104(14) & -0.00021(17) \\
S1 & 0.0091(7) & 0.0080(7) & 0.0098(7) & 0 & -0.0005(5) & 0 \\
S2 & 0.0097(7) & 0.0080(7) & 0.0100(7) & 0 & 0.0004(5) & 0 \\ \hline
\end{tabular}
\label{T3}
\end{table}
\end{footnotesize}

\end{document}